\documentclass[12pt, a4paper]{article}
\usepackage[utf8]{inputenc}
\usepackage{physics} 
\usepackage{amsmath}
\usepackage{amsfonts}
\usepackage{amssymb}
\usepackage{graphicx}
\usepackage{authblk}
\usepackage{xcolor}
\usepackage{subcaption}
\usepackage{cite}
\DeclareUnicodeCharacter{0327}{\c{}}
\usepackage[colorlinks=true,linkcolor=blue,citecolor=blue,urlcolor=blue]{hyperref}

\newcommand {\bg}{\bar \gamma}
\newcommand {\bp}{\bar \psi}

\usepackage[left = 1.5cm, right = 1.5cm, top = 1cm, bottom = 2cm]{geometry}
\begin{document}

\begin{center}

{\large {\bf Testing an anisotropic spinor field–based Modified Chaplygin Gas model in Kantowski--Sachs spacetime with observational constraints
}}
\vskip .5cm
\textbf{Mahendra Goray$^{a}$ Bijan Saha$^{b,c}$}
\end{center}
\vskip0.1cm
$^a$Department of Physics, S.P. College, Sido Kanhu Murmu University, Dumka-814101, Jharkhand, India\\
$^b$Laboratory of Information Technologies, Joint Institute for Nuclear Research,
141980 Dubna, Moscow region, Russia\\
$^c$Peoples' Friendship University of Russia (RUDN University), 6 Miklukho-Maklaya Street, Moscow, Russian Federation\\

\textbf{E-mail:} goraymahendra92@gmail.com \textbf{(M.Goray)}, bijan@jinr.ru \textbf{(B.Saha)}

\begin{abstract}
We investigate a cosmological model based on a massless nonlinear spinor field coupled to a Modified Chaplygin Gas (MCG) in the Kantowski--Sachs spacetime, aiming to probe anisotropies and unified dark sector dynamics. The model parameters are constrained using recent observational data, including Pantheon+, cosmic chronometers, DESI DR2, and CMB distance priors, via a Markov Chain Monte Carlo analysis. We find $H_0 \sim 67$--$68~\mathrm{km\,s^{-1}\,Mpc^{-1}}$, while the shear parameter is consistent with zero, indicating an effectively isotropic Universe at late times. The model reproduces late-time cosmic acceleration with a present-day deceleration parameter $q_0 \sim -0.49$, and provides a good fit to the data, yielding a lower minimum $\chi^2$ than $\Lambda$CDM, and is favored by the Akaike Information Criterion. Overall, the spinor field MCG model in Kantowski--Sachs spacetime offers a viable framework that naturally incorporates anisotropy and a unified description of dark matter and dark energy, consistent with current observations.
\end{abstract}

\section*{I. Introduction}
The standard concordance model of cosmology, or simply the standard model of cosmology (SMC), also known as the model with a cosmological constant and Cold Dark Matter and commonly referred to as the $\Lambda$CDM model, provides a remarkably successful description of the large-scale structure and evolution of the Universe. This framework is based on three fundamental assumptions: (i) the validity of general relativity on cosmological scales, (ii) the applicability of the standard model of particle physics at microscopic scales, and (iii) the cosmological principle, which states that the Universe is spatially homogeneous and isotropic on large scales \cite{Heavenscon, Weinberg2008, Dodelson2003}. Within this model, the Universe evolves from an initial hot and dense state and is currently composed of approximately $5\%$ baryonic matter, $27\%$ dark matter, and $68\%$ dark energy \cite{Planck2018, PeeblesRatra2003, Heavens}.

The $\Lambda$CDM model successfully fits a wide range of observations, including type Ia supernovae \cite{Riess1998, Perlmutter1999, Pantheon2018, PantheonPlus2022}, cosmic microwave background (CMB) anisotropies\cite{Planck2018, Planck2013}, large-scale structure formation \cite{Tegmark2004, Cole2005}, weak gravitational lensing \cite{Bartelmann2001, Kilbinger2015}, and baryon acoustic oscillations \cite{eBOSS2021, DESI2024, DESI2023, DESIDR1, DESIDR2}. In particular, it provides a consistent description of the observed late-time accelerated expansion of the Universe. However, despite its observational success, the model faces several theoretical and observational challenges, such as the cosmological constant fine-tuning problem and the persistent Hubble tension \cite{Weinberg1989, Carroll2001, Riess2019, Verde2019}. Moreover, the dark energy component in $\Lambda$CDM is characterized by a constant equation of state parameter, $w_\Lambda = -1$, reflecting its non-dynamical nature \cite{Weinberg2008, PeeblesRatra2003, Copeland2006}. These limitations motivate the exploration of alternative cosmological models that allow for dynamical dark energy and possible deviations from the standard assumptions.

Since the cosmological principle constitutes a fundamental assumption of the $\Lambda$CDM model, any deviation from large-scale homogeneity and isotropy may be interpreted as a modification of the standard framework. Observational evidence, particularly from the cosmic microwave background and large-scale structure, strongly supports the isotropy and homogeneity of the Universe at sufficiently large scales, thereby justifying the widespread use of the Friedmann–Lemaître–Robertson–Walker (FLRW) metric in cosmological studies. However, both theoretical considerations and observational indications suggest that the early Universe may have undergone an anisotropic phase, which subsequently evolved toward isotropy \cite{Misner1968, Ellis1969, SahaPRD2001, PlanckIsotropy2016, Saadeh2016}. Such possibilities motivate the study of more general cosmological models that relax the assumption of isotropy. In this context, anisotropic space-times such as the Kantowski–Sachs spacetime and Bianchi models provide a natural generalization of the standard framework \cite{Kantowski1966, Ellis1969, Ryan1975, Saha2018}. While Bianchi-type models have been extensively investigated in earlier works, the present study focuses on the Kantowski–Sachs geometry as a viable anisotropic extension of the standard cosmological model.

Spinor fields have attracted considerable attention in cosmology and astrophysics due to their ability to effectively model a wide range of matter sources, including perfect fluids and dark energy components. In particular, the structure of the spinor affine connection in the covariant derivative gives rise to nontrivial non-diagonal components in the energy–momentum tensor. These components impose significant constraints on both the spacetime geometry and the spinor field configuration, thereby enriching the dynamical behavior of the system. The role of spinor fields in the evolution of the Universe has been extensively studied in both isotropic and anisotropic cosmological settings \cite{SahaPRD2001, goray2025, goray2026, Saha2015APSS, Fabri, Pop, Saha2018}. In this work, we extend such investigations to a cosmological model based on the Kantowski–Sachs spacetime, which is spatially homogeneous and rotation-free. Owing to the presence of shear, this geometry is generally regarded as anisotropic, although certain studies suggest the possibility of isotropization under specific conditions \cite{Torrence}. Recently, Kantowski–Sachs models have been explored in the context of various modified theories of gravity \cite{Krasnov, Vinutha}. The dynamics of spinor fields in Kantowski–Sachs (KS) spacetime have also been investigated in recent studies \cite{Saha2025}, further motivating the present analysis.

A variety of dynamical dark energy models have been proposed in the literature, including quintessence and Chaplygin gas–based scenarios, in order to describe the accelerated expansion of the Universe within both isotropic and anisotropic cosmological frameworks \cite{Ratra1988, Caldwell1998, Kamenshchik2001, Bento2002, Copeland2006}. Among these, the Modified Chaplygin Gas (MCG) has attracted significant attention due to its ability to provide a unified description of dark matter and dark energy through a single equation of state. This feature eliminates the need to treat these components separately, offering a more economical and flexible framework for cosmological modeling \cite{Benaoum2002, Debnath2004, Lu2008}. In this work, we adopt the Modified Chaplygin Gas as the dark energy candidate and investigate its dynamics in the presence of a spinor field within the KS geometry. This combined framework enables us to explore the interplay between anisotropy, fermionic fields, and unified dark fluid models in shaping the expansion history of the Universe.

Moreover, within the framework of the Kantowski–Sachs spacetime, we construct an MCG model driven by a massless nonlinear spinor field and constrain its parameters using the Markov Chain Monte Carlo (MCMC) method. The parameter set includes the present Hubble parameter $H_0$, the curvature parameter $\Omega_{k0}$, the present shear parameter $\Delta_0$, along with the MCG parameters $A$, $\alpha$, and $W$. To obtain robust constraints, we employ recent cosmological observations, including the Type Ia Supernovae distance modulus data from the Pantheon+ compilation, Hubble parameter measurements from cosmic chronometers, baryon acoustic oscillation data from the DESI Data Release 2 (DR2), and cosmic microwave background observations. For a comprehensive statistical comparison, we also analyze both curved and flat $\Lambda$CDM models using the same data sets.

The manuscript is organized as follows. In Section II, we present the mathematical formulation of the model within the Kantowski–Sachs spacetime. Section III is devoted to the construction of the Modified Chaplygin Gas model. In Section IV, we describe the observational data sets and the Markov Chain Monte Carlo (MCMC) methodology used to constrain the model parameters. Section V presents and analyzes the numerical results. Finally, the discussion and conclusions are summarized in Section VI.

\section*{II. Mathematical background}

We consider the spinor field Lagrangian given by
    
	\begin{align}
		L_{\rm sp} = \frac{\imath}{2} \biggl[\bp \gamma^{\mu} \nabla_{\mu}
		\psi- \nabla_{\mu} \bar \psi \gamma^{\mu} \psi \biggr] - m_{\rm sp}
		\bp \psi - \lambda F(S), \label{lspin}
	\end{align}
	where $\lambda$ is the self-coupling constant and $F$ is the spinor field nonlinearity, $S = \bp \psi$. 
	
	The space-time is given by the Kantowski-Sachs metric 
	
	\begin{align}
		ds^2 = dt^2 - a_1^2 dr^2 - a_2^2 \left(d\theta^2 +
		\sin^2 \theta d\phi^2\right), \label{KS}
	\end{align}
	with $a_1,\, a_2$  being the functions of time only. As one sees, in the case of $a_2 = a_1$, the Kantowski-Sachs metric dully transforms into flat FLRW space-time. 
	
The spinor field equation in this case is 
\begin{subequations}
  \label{SFE1} 
\begin{align}\dot \psi +  \frac{1}{2} \left(H_1 + 2 H_2\right)\psi +\imath m_{\rm sp} \bg^0 \psi +\imath F_S \bg^0 \psi  & = 0, \label{speq1r} \\
\dot \bp +  \frac{1}{2} \left(H_1 + 2 H_2\right)\bp -\imath  m_{\rm sp} \bp \bg^0 - \imath F_S \bp \bg^0  & = 0, \label{speq2r}
\end{align}
\end{subequations}
where $F_S = dF/dS$, $H_1 = \dot a_1 / a_1$ and $H_2 = \dot a_2 /a_2.$ 
For $S$ in this case we have 
\begin{align}
\dot S + \left(H_1 + 2 H_2\right) S &= 0, \label{S} 
\end{align}
with the solution 
\begin{align}
S &= \frac{S_0}{V}, \quad S_0 = {\rm const.} \label{Sdef} 
\end{align}
where we define volume scale as
\begin{align}
V &= a_1 a_2^2. \label{Vdef} 
\end{align}
The EMT in this case 
\begin{subequations} 
\label{emtnd}
\begin{align}
T_0^0 &= m_{\rm sp} S + \lambda F = \rho , \label{T00} \\
T_1^1 &= T_2^2 = T_3^3 = \lambda \left(F - 2 K F_K\right) = - p. \label{Tii}\\
T_2^1 &=  \frac{1}{4}\frac{a_2}{a_1} \left(\frac{\dot a_1}{a_1} - \frac{\dot a_2}{a_2}\right) A^3, \label{emt12} \\ 
T_3^1 &= \frac{1}{4}\frac{a_2}{a_1} \sin \theta \left(\frac{\dot a_1}{a_1} - \frac{\dot a_2}{a_2}\right) A^2. \label{emt31}
\end{align}
\end{subequations} 
where $A^\mu = \bp \gamma^5 \gamma^\mu \psi.$ Since Einstein tensor for Kantowski-Sachs model possesses only diagonal components, this leads to either $A^2 = A^3 = 0$ or $(a_1^{-1}\,\,\dot a_1 - a_2^{-1}\,\,\dot a_2) = 0.$

The diagonal components of Einstein equations 
	\begin{subequations} 
		\label{EED} 
		\begin{align}
			2 \dot H_2 + 3 H_2^2  + \frac{1}{a_2^2}
			&= - 8\pi G p, \label{E11}\\
			\dot H_1  + \dot H_2 +   H_1^2 + H_2^2 + H_1 H_2 &= - 8\pi G p, 
			\label{E22}\\ 
			2 H_1 H_2 
			+  H_2^2 + \frac{1}{a_2^2} &= 8\pi G \rho. \label{E00}
		\end{align}
	\end{subequations} 
    Let us define the mean scale factor as 
    \begin{equation}
        a = \left(a_1 a_2^2\right)^{1/3}, \label{meana}
            \end{equation}
that leads to the mean Hubble parameter $H$:
\begin{equation}
H = \frac{\dot a}{a} = \frac{1}{3}\left(\frac{\dot a_1}{a_1} + 2\frac{\dot a_2}{a_2}\right) = \frac{1}{3}(H_1 + 2H_2). \label{mean_H}
\end{equation}
From \eqref{mean_H} one can dully find
\begin{equation}
H^2 = \frac{1}{9}\left(H_1^2 + 4H_2^2 + 4H_1 H_2\right). \label{mean_H^2}
\end{equation}
Let us also define a new parameter 
\begin{equation}
    \Delta = H_1 - H_2. \label{delta}
\end{equation}
In view of \eqref{mean_H} and \eqref{delta} the directional Hubble parameters can be written as
\begin{subequations}
    \begin{align}
        H_1 =H+\dfrac{2}{3}\Delta\\
        H_2 =H-\dfrac{1}{3}\Delta.
    \end{align}
\end{subequations}
For the shear scalar $\sigma^2 = \left(\sigma_{\mu\nu} \sigma^{\mu\nu}\right)/2$ in this case we find
\begin{equation}
\sigma^2 = \frac{1}{3}(H_1 - H_2)^2, \label{sigma^2}
\end{equation}
that gives 
\begin{align}
    \Delta^{2} = 3 \sigma^2. \label{delta_sigma}
\end{align}
Equation \eqref{mean_H^2} in view of \eqref{sigma^2} can be presented as
\begin{equation}
H_2^2 + 2H_1 H_2 = 3H^2 - \sigma^2. \label{H2_sigma2}
\end{equation}
On account of~\eqref{H2_sigma2} Equation~\eqref{E00} can be written as 
\begin{equation}
3H^2 = 8\pi G \rho - \frac{1}{a_2^2} + \sigma^2. \label{H2sigma2k2}
\end{equation}

\section*{III. The Model}
    
We consider the case with modified Chaplygin gas (MCG). The MCG was proposed in order to unify both dark matter and dark energy, and given by the equation of state (EOS) 
	\begin{align}
		p &=  W \rho - \frac{A}{\rho^ \alpha}, \quad W >0, \quad  A >0, \quad 0 \le \alpha \le 1,  
		\label{modchap0}  
	\end{align} 
Then inserting $\rho = \lambda F(K)$ and $p = \lambda \left(2 K F_K - F(K)\right)$ into \eqref{modchap0} for massless spinor field $m_{\rm sp}=0$, 
we find the corresponding spinor field nonlinearity: 
	\begin{align}
		F(K) &= \left[\frac{A}{1+W} + \lambda_1
		K^{(1+\alpha)(1+W)/2}\right]^{1/(1+\alpha)}. \label{modchap0F}
	\end{align} 
In the above expressions $\lambda_1$ is the constant of integration. 
Corresponding energy density of the modified Chaplygin gas	
\begin{equation}
    \rho = \lambda \left[\frac{A}{1+W} + \lambda_1
		K^{(1+\alpha)(1+W)/2}\right]^{1/(1+\alpha)}. \label{rho_mcg}
\end{equation}

The evolution equations for the spinor invariants, leading to
\begin{align}
K = \frac{K_0}{a^{6}} = K_0 (1+z)^6, \qquad K_0 = \text{const}.
\label{Kdef0_new}
\end{align}
This relation holds for a massless spinor field when
$K = \{J,\, I \pm J\}$, whereas for $K = I$ it is valid for both
massive and massless cases. Where, $1+z = a_0/a$ and $a_0$ is the present scale factor, taken to be unity: $a_0 = 1$.
On account of \eqref{Kdef0_new}  energy density and pressure of the system, we rewrite in terms of redshift: 

\begin{subequations}
	\begin{align}
		\rho &=  \lambda \left[\frac{A}{1+W} + \lambda_1
		(1+z)^{3(1+\alpha)(1+W)}\right]^{1/(1+\alpha)} ,  \label{mchapsped}\\
		p &=  \lambda \left[\lambda_1 W
		(1+z)^{3(1+\alpha)(1+W)} - \frac{A}{1+W}\right] \left[\frac{A}{1+W} + \lambda_1
		(1+z)^{3(1+\alpha)(1+W)}\right]^{-\alpha/(1+\alpha)}
		. \label{modchapp} 
	\end{align}
\end{subequations}

On account of \eqref{delta_sigma} we rewrite $H^2$ in Equation~\eqref{H2sigma2k2},
\begin{equation}
H^{2} = \frac{8\pi G \rho}{3} + \frac{1}{9}\Delta^{2} - \frac{1}{3a_2^{2}}. \label{HDelta}
\end{equation}

Hence the normalized Hubble parameter becomes
\begin{equation}
E^{2}(z) = \frac{H^{2}}{H_0^{2}}
= \frac{8\pi G \rho}{3H_0^{2}}
+ \frac{\Delta^{2}}{9H_0^{2}}
- \frac{1}{3H_0^{2}a_2^{2}} .
\end{equation}

In view of \eqref{meana} we define
\begin{equation}
a_2 = a e^{-\beta}, \qquad
a_1 = a e^{2\beta},
\end{equation}

Hence,
\begin{equation}
H_1 = \frac{\dot a_1}{a_1} = H + 2\dot{\beta},
\end{equation}

\begin{equation}
H_2 = \frac{\dot a_2}{a_2} = H - \dot{\beta}.
\end{equation}

Note that in this case, the mean Hubble parameter \eqref{mean_H} remains unaltered. Moreover, from \eqref{delta} we obtain  
\begin{equation}
\Delta = H_1 - H_2 = 3\dot{\beta}.
\end{equation}

Now into account that $a$ is related to red-shift $z$ as 
\begin{equation}
    z + 1 = \frac{1}{a},\label{R-S}
    \end{equation}
    we find
\begin{equation}
\frac{1}{a_2^{2}}
=
\frac{1}{a^{2}}e^{2\beta}
=
(1+z)^{2}e^{2\beta(z)} .
\end{equation}

Now the dimensionless Hubble parameter $E(z)$ in the Kantowski--Sachs spacetime is given by,
\begin{equation}
E^2(z) =
\frac{8\pi G \rho(z)}{3H_0^2} + \frac{\Delta^{2}}{9H_0^{2}}
- \frac{1}{3H_0^2}(1+z)^2 e^{2\beta(z)} .
\end{equation}

\begin{align}
E^2(z)
&=
\frac{8\pi G \rho_{\mathrm{MCG},0}}{3H_0^2}
\frac{\rho_{\mathrm{MCG}}(z)}{\rho_{\mathrm{MCG},0}} + \frac{\Delta^{2}}{9H_0^{2}}
- \frac{1}{3H_0^2}(1+z)^2 e^{2\beta(z)}
\end{align}

\begin{equation}
E^2(z) = \Omega_{\text{MCG,0}} \, \tilde{\rho}_{\text{MCG}}(z)
+ \frac{\Delta^2(z)}{9H_0^2}
+ \Omega_{k,0} (1+z)^2 e^{2\beta(z)},
\end{equation}

At \(z=0\), $E^2(0)=1$, and $\beta(0)\xrightarrow{}0$ as present day epoch is isotropic, with the effective density parameter for MCG is
\begin{equation}
\Omega_{\text{MCG},0} = 1 - \Omega_{k,0} - \frac{\Delta_0^2}{9H_0^2}.
\end{equation}
Here, $H_0$ is the present Hubble parameter, $\Omega_{k,0}$ is the present curvature density parameter, $\Delta(z)$ is the anisotropy parameter, and $\beta(z)$ characterizes the anisotropic expansion.
The evolution equations for the Kantowski--Sachs model are given by the coupled differential equations:

\begin{equation}
\frac{d\beta}{dz} = -\frac{\Delta(z)}{3(1+z)H(z)},
\end{equation}

\begin{equation}
\frac{d\Delta}{dz} = \frac{3\Delta(z)}{1+z} - \frac{(1+z)e^{2\beta(z)}}{H(z)},
\end{equation}

where the Hubble parameter is related to $E(z)$ as $H(z) = H_0 \, E(z)$.

The effective equation of state paremeter for spinor field MCG dark energy $w(z) = \frac{p(z)}{\varepsilon(z)}$, and the deceleration parameter $q(z)=-\frac{\ddot a}{a}\frac{1}{H^{2}}$ for this MCG model can be constructed as 
 \begin{align}
     q(z)= -1+(1+z)\frac{1}{E}\frac{dE}{dz}. \label{Decpar}
 \end{align}

\section*{IV. Method and Data sets}
To constrain the spinor field MCG model in the Kantowski–Sachs framework, we employed a combination of recent cosmological data sets. The model uses a joint likelihood built from Cosmic Chronometers (CC), type~Ia supernovae (Pantheon+), Baryon Acoustic Oscillation (BAO) measurements, and cosmic microwave background (CMB) measurements.

\begin{itemize}
\item  {\textbf{Cosmic Chronometers (CC):}}
  We used 31 Hubble parameter measurements from cosmic chronometers (CC) \cite{Stern2010, Moresco2012, Moresco2016, Zhang2014, Simon2005, Delubac2015}, which provide direct measurements of the Hubble parameter $H(z)$ in between $0<z<2$. The CC chi-square is
\begin{equation}
\chi^2_{\rm CC}
=\sum_i
\frac{\left[H_{\rm obs}(z_i) - H_{\rm th}(z_i)\right]^2}
{\sigma_{H,i}^2}, 
\end{equation}
where $H_{\rm obs}(z_i)$, $H_{\rm th}(z_i)$, and $\sigma_{H,i}$ denote the observed Hubble parameter, the theoretical prediction, and the corresponding uncertainty at redshift $z_i$, respectively.

\item {\textbf{Pantheon+ supernovae (PP):}}
We include 1588 Pantheon+ Type Ia supernovae with their full covariance from the SNIa dataset \cite{PantheonPlus2022}. We adopt the Pantheon+ sample with a redshift cut of $z > 0.01$ to minimize systematic effects from peculiar velocities, and we use the uncalibrated distances (i.e., without imposing a prior on $H_0$ from SH0ES Collaboration). The Pantheon+ compilation provides measurements of the distance modulus $\mu_i$ at redshifts $z_i$ together with the full statistical and systematic covariance matrix $C$. The theoretical distance modulus is
\begin{equation}
\mu_{\rm th}(z)=5\log_{10}D_L(z)+25,
\end{equation}
where the luminosity distance is
\begin{equation}
D_L(z)=(1+z)D(z),
\end{equation}
and the comoving distance $D(z)$ satisfies
\begin{equation}
D(z) = \frac{c}{H_0} \int_0^z \frac{dz'}{E(z')}
\end{equation}
with $E(z)=H(z)/H_0$. The absolute magnitude is analytically marginalized. Defining $\Delta\mu=\mu_{\rm obs}-\mu_{\rm th}$, the supernova chi-square is 
\begin{equation}
\chi^2_{\rm SN}
=\Delta\mu^{T}C^{-1}\Delta\mu
-\frac{(\mathbf{1}^{T}C^{-1}\Delta\mu)^2}
{\mathbf{1}^{T}C^{-1}\mathbf{1}} .
\end{equation}

\item  {\textbf{ DESI DR2:}} We use the latest 13 DESI BAO compilation including DR2 constraints at redshifts
$z=\{0.295,0.510,0.706,0.934,1.321,1.484, 2.33\}$, including the observables $D_V/r_d$, $D_M/r_d$, and $D_H/r_d$, where the sound horizon is fixed to $r_d = 147.05~{\rm Mpc}$ \cite{DESI_DR2_BAO_2025, DESI_DR2_Lya_2025}. For simplicity, we adopt the Gaussian approximation provided in the Cobaya BAO compilation and treat the measurements as independent, without including the full covariance matrix.

The BAO distances are
\begin{equation}
D_H(z)=\frac{c}{H(z)}, \qquad
D_M(z)=
\begin{cases}
\frac{c}{H_0\sqrt{\Omega_k}} \sinh(\sqrt{\Omega_k}\chi), & \Omega_k>0 \\
\frac{c}{H_0}\chi, & \Omega_k=0 \\
\frac{c}{H_0\sqrt{-\Omega_k}} \sin(\sqrt{-\Omega_k}\chi), & \Omega_k<0
\end{cases}
\end{equation}

\begin{equation}
D_V(z)=\left[zD_M^2(z)D_H(z)\right]^{1/3},
\end{equation}
where
\begin{equation}
\chi(z)=\frac{c}{H_0}\int_0^z \frac{dz'}{E(z')}.
\end{equation}
The BAO chi-square is
\begin{equation}
\chi^2_{\rm BAO}
=\sum_i
\frac{\left[\mathcal{O}_i-\mathcal{O}_{\rm th}(z_i)\right]^2}
{\sigma_i^2}.
\end{equation}

\item {\textbf{Cosmic Microwave Background Distance Priors (CMB):}} To incorporate early-Universe constraints, we use the Planck 2018 distance priors \cite{Planck2018}, namely the shift parameter $R$ and the acoustic scale $\ell_A$.

The shift parameter is
\begin{equation}
R=\sqrt{\Omega_m} H_0
\frac{D_M(z_*)}{c},
\end{equation}
and the acoustic scale is
\begin{equation}
\ell_A=\pi\frac{D_M(z_*)}{r_d},
\end{equation}
where $z_*=1089.92$ is the redshift of recombination.
The observed values are
\begin{equation}
R_{\rm obs}=1.74963, 
\qquad
\ell_{A,{\rm obs}}=301.80845,
\end{equation}
with covariance
\begin{equation}
C_{\rm CMB}=
\begin{pmatrix}
\sigma_R^2 & \rho\sigma_R\sigma_{\ell_A} \\
\rho\sigma_R\sigma_{\ell_A} & \sigma_{\ell_A}^2
\end{pmatrix},
\end{equation}
where $\sigma_R=0.0040$, $\sigma_{\ell_A}=0.0900$, and $\rho=0.53$.
The CMB chi-square is
\begin{equation}
\chi^2_{\rm CMB}
=\Delta\mathbf{Y}^T
C_{\rm CMB}^{-1}
\Delta\mathbf{Y},
\end{equation}
where $\Delta\mathbf{Y}=(R-R_{\rm obs},\,\ell_A-\ell_{A,{\rm obs}})$.

For the CMB likelihood, we approximate the effective matter density at early times as
\begin{equation}
\Omega_m^{\rm eff} = \frac{1 - \Omega_k - \Delta_0^2/(9H_0^2)}{1 + A/(1+W)},
\end{equation}
which ensures consistency with the matter-dominated regime at high redshift.

\end{itemize}

To parametrize the model parameters, we used the \texttt{emcee} Python library for performing the MCMC technique in Python 3 ($\it{ipykernel}$) using three different combinations of data sets \textbf{(i.)} PP + CC \textbf{(ii.)} PP + CC + DESI DR2 \textbf{(iii.)} PP + CC + DESI DR2 + CMB. Along with this, we also constrain the both curved and flat $\Lambda$CDM models using the same combined data sets in order to provide a direct comparison with the spinor field MCG model.  
The total likelihood is written as
\begin{equation}
\mathcal{L}
=\mathcal{L}_{\rm CC}\,
\mathcal{L}_{\rm SN}\,
\mathcal{L}_{\rm BAO}\,
\mathcal{L}_{\rm CMB}.
\end{equation}
The full log-likelihood used in the MCMC analysis is
\begin{equation}
\ln\mathcal{L}
=-\frac12\left(
\chi^2_{\rm CC}
+\chi^2_{\rm SN}
+\chi^2_{\rm BAO}
+\chi^2_{\rm CMB}
\right).
\end{equation}

We explore the six-dimensional parameter space; paramas~$ = H_0, \Omega_{k0}, \Delta_{0}, A, \alpha, W$ where $A$, $\alpha$ are model parameters. However, in our analysis, we took the priors of the present Hubble parameter in $50 < H_0 < 80$, present curvature density parameter and shear parameter as $-0.02<\Omega_{k0}<0.02$ and $-3<\Omega_0 <3$ respectively. The
other priors to the model parameters: $0 < A < 5$, $0 < \alpha < 1$, and $-2 < W < 0$. For MCMC simulation, we set the value of the other two constants of the MCG model as  $\lambda_{1}=1.0$ and $\lambda =1.0$. The sound horizon scale is fixed to $r_d=147.05$ Mpc for consistency with legacy BAO analyses. To run the MCMC, we set walkers $= 48$ and steps $=3000$.

\begin{table*}[htbp]
\centering
\caption{Summary of parameter constraints at the $1\sigma$ confidence level (median $\pm 1\sigma$) for all considered cosmological models across different combinations of observational data sets.}
\begin{tabular}{llccc}
\hline\hline
Model & Parameter & PP+CC & PP+CC+DESI DR2 & PP+CC+DESI DR2+CMB \\
\hline

{KS+Spinor MCG}
& $H_0$ & $67.64^{+1.79}_{-1.70}$ & $68.37^{+0.59}_{-0.57}$ & $67.74^{+0.50}_{-0.55}$ \\[0.1cm]
& $\Omega_{k0}$ & $0.071^{+0.095}_{-0.148}$ & $0.098^{+0.044}_{-0.046}$ & $0.008^{+0.018}_{-0.017}$ \\[0.1cm]
& $\Delta_0$ & $0.160^{+1.97}_{-2.13}$ & $-0.205^{+2.36}_{-2.12}$ & $-0.006^{+2.32}_{-2.28}$ \\[0.1cm]
& $A$ & $1.16^{+0.64}_{-0.46}$ & $2.20^{+0.33}_{-0.30}$ & $1.65^{+0.17}_{-0.17}$ \\[0.1cm]
& $\alpha$ & $0.412^{+0.373}_{-0.287}$ & $0.120^{+0.137}_{-0.086}$ & $0.086^{+0.134}_{-0.065}$ \\[0.1cm]
& $W$ & $-0.179^{+0.094}_{-0.082}$ & $-0.031^{+0.021}_{-0.029}$ & $-0.066^{+0.021}_{-0.028}$ \\[0.1cm]

\hline

{Curved $\Lambda$CDM}
& $H_0$ & $67.23^{+1.75}_{-1.71}$ & $68.26^{+0.47}_{-0.48}$ & $68.13^{+0.48}_{-0.48}$ \\[0.1cm]
& $\Omega_{m0}$ & $0.303^{+0.067}_{-0.045}$ & $0.282^{+0.015}_{-0.014}$ & $0.328^{+0.006}_{-0.006}$ \\[0.1cm]
& $\Omega_{k0}$ & $0.084^{+0.082}_{-0.127}$ & $0.089^{+0.038}_{-0.036}$ & $-0.031^{+0.008}_{-0.007}$ \\[0.1cm]

\hline

{Flat $\Lambda$CDM}
& $H_0$ & $67.08^{+1.70}_{-1.68}$ & $68.39^{+0.47}_{-0.47}$ & $66.58^{+0.23}_{-0.23}$ \\[0.1cm]
& $\Omega_{m0}$ & $0.345^{+0.018}_{-0.017}$ & $0.312^{+0.009}_{-0.008}$ & $0.351^{+0.001}_{-0.001}$ \\[0.1cm]

\hline\hline
\end{tabular}
\label{tab:parameter_constraints}
\end{table*}

\begin{figure}[!ht]
\centering
\includegraphics[width=0.8\textwidth]{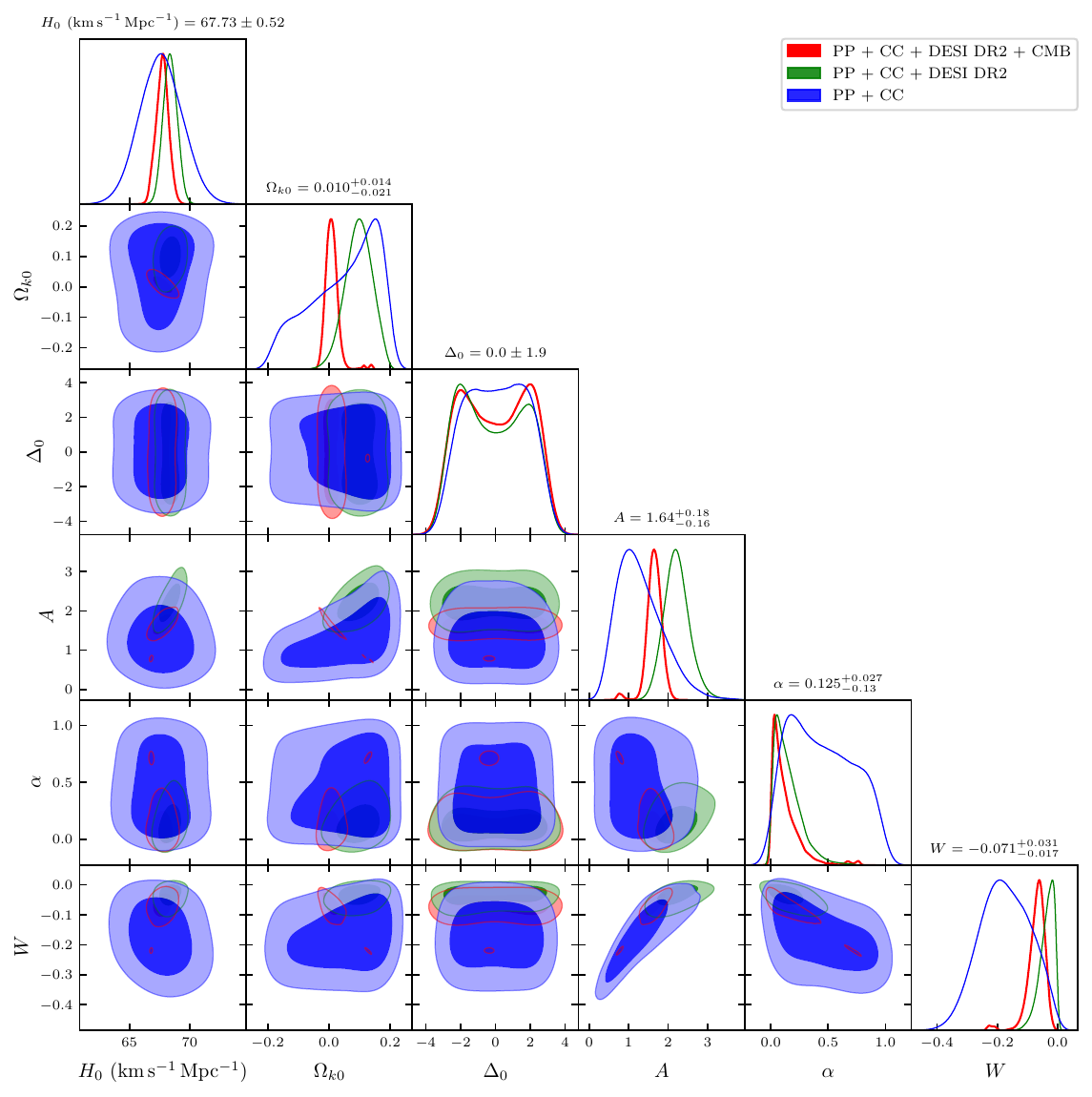}
\caption{Marginalized posterior distributions of the KS+Spinor MCG model parameters for the combined data sets: PP+CC, PP+CC+DESI DR2, and PP+CC+DESI DR2+CMB.}
\label{Fig:mcg1}
\end{figure}

\begin{figure}[!ht]
\centering
\includegraphics[width=0.8\textwidth]{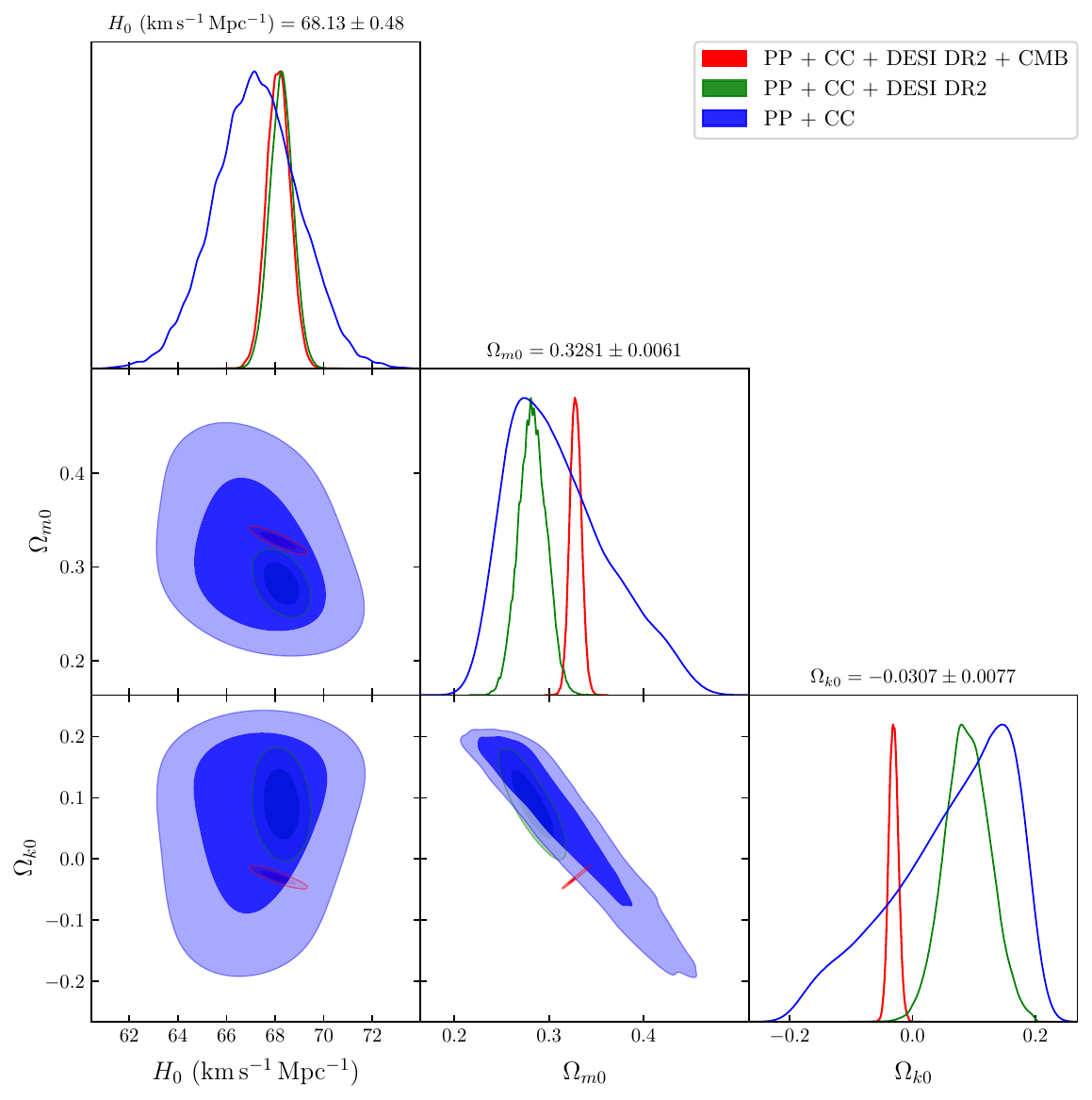}
\caption{Marginalized posterior distributions of the curved $\Lambda$CDM model parameters for the combined data sets: PP+CC, PP+CC+DESI DR2, and PP+CC+DESI DR2+CMB.}
\label{Fig:mcg2}
\end{figure}

\begin{figure}[!ht]
\centering
\includegraphics[width=0.8\textwidth]{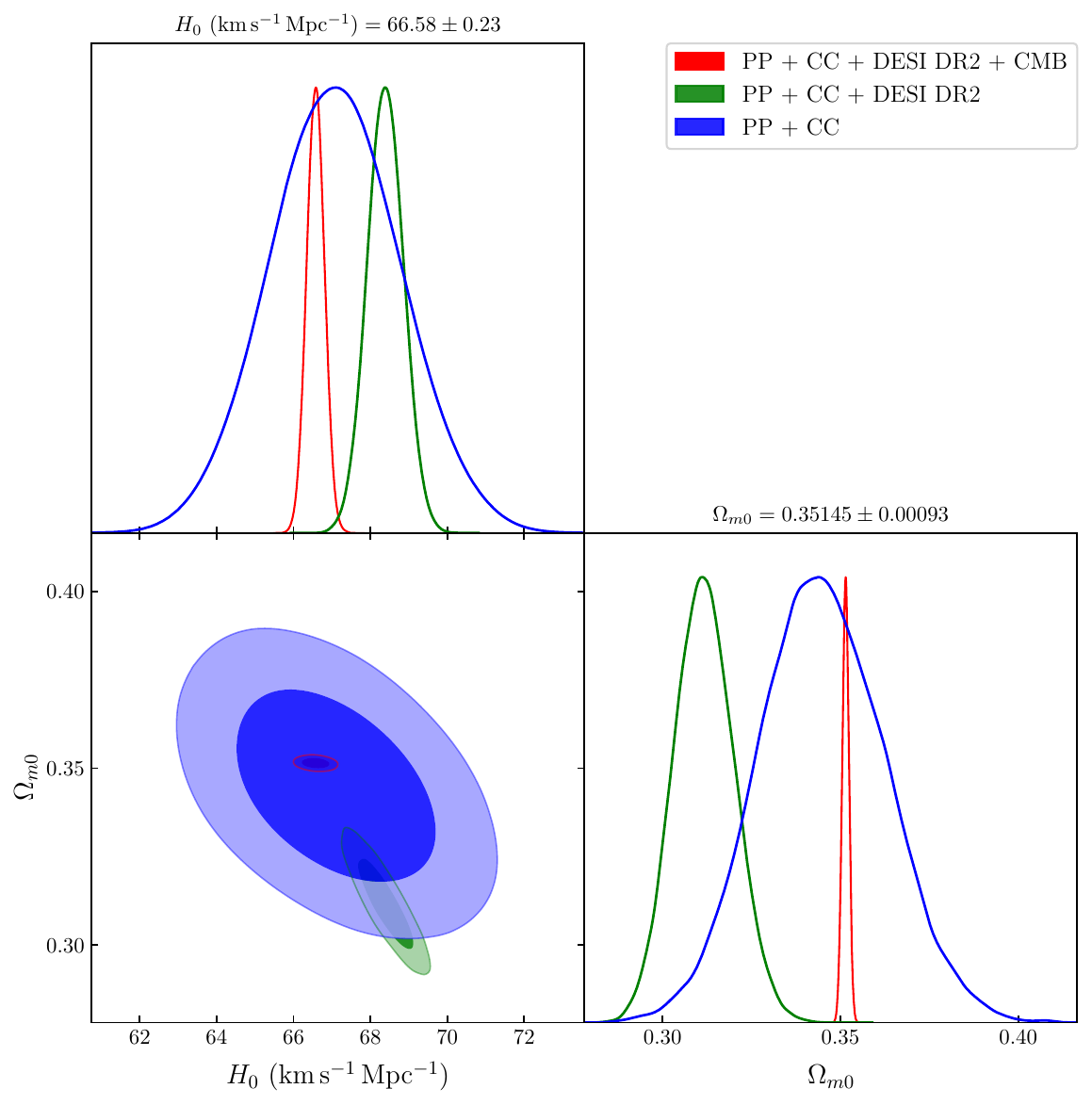}
\caption{Marginalized posterior distributions of the flat $\Lambda$CDM model parameters for the combined data sets: PP+CC, PP+CC+DESI DR2, and PP+CC+DESI DR2+CMB.}
\label{Fig:mcg3}
\end{figure}

\section*{V. Results and Discussion}
We present the observational constraints on the spinor field Modified Chaplygin Gas (MCG) model in the Kantowski--Sachs (KS) spacetime (KS+Spinor MCG), along with the curved and flat $\Lambda$CDM models, using three combinations of data sets: (i) PP + CC, (ii) PP + CC + DESI DR2, and (iii) PP + CC + DESI DR2 + CMB. The results are summarized in Table~\ref{tab:parameter_constraints}. Corresponding the marginalized posterior distributions and confidence contours ($\pm 1\sigma$) are shown in Figs.~\ref{Fig:mcg1}, \ref{Fig:mcg2}, and \ref{Fig:mcg3} which provide further insight into parameter correlations.

The normalized best-fit values of the parameters from the MCMC analysis are $[H_0, \Omega_{k0}, \Delta_0, A, \alpha, W] = [67.667, 0.016, -2.934, 1.599, 0.009, -0.057]$ for the KS+Spinor MCG model. The Hubble parameter $H_0$ remains remarkably stable across all data combinations, with values centered around $H_0 \sim 67$--$68~\mathrm{km\,s^{-1}\,Mpc^{-1}}$. The inclusion of DESI DR2 and CMB data significantly reduces the uncertainties, yielding $H_0 = 67.74^{+0.50}_{-0.55}$ for the full data set. This value is consistent with Planck CMB measurements and indicates that the KS+Spinor MCG model naturally aligns with early-Universe constraints. In comparison, the curved and flat $\Lambda$CDM models yield $H_0 = 68.13 \pm 0.48$ and $H_0 = 66.58 \pm 0.23$, respectively.

The curvature parameter $\Omega_{k0}$ shows an interesting evolution. For the PP+CC data, a mildly open Universe is preferred ($\Omega_{k0} \sim 0.07$), while the addition of DESI DR2 tightens the constraint. With the inclusion of CMB data, the curvature parameter shifts toward $\Omega_{k0} = 0.008^{+0.018}_{-0.017}$, which is fully consistent with a spatially flat Universe within $1\sigma$ in the current epoch.
While for the $\Lambda$CDM model it approaches $\Omega_{k0} =  -0.031^{+0.008}_{-0.007}$ This demonstrates the strong constraining power of CMB observations. 

The shear parameter $\Delta_0$ is found to be very weakly constrained across all data sets, with large uncertainties encompassing zero. For the full data combination, $\Delta_0 = -0.006^{+2.32}_{-2.28}$, indicating that the present Universe is effectively isotropic. This result is consistent with the expectation that any primordial anisotropy decays over cosmic evolution.

The Modified Chaplygin Gas parameters exhibit significant improvement with additional data. The parameter $A$  shifts from $A \sim 1.16$ (PP+CC) to $A \sim 1.65$ (full data), while its uncertainty decreases substantially. Similarly, the parameter $\alpha$ becomes tightly constrained toward small values, $\alpha \approx 0.086$, indicating a mild deviation from the standard Chaplygin gas behavior. The intrinsic equation of state parameter $W$ remains negative and approaches $W \approx -0.066$, supporting a dark energy–like behavior at late times.

In Table~\ref{tab:parameter_constraints}, we present a statistical comparison between the KS+Spinor MCG model and both $\Lambda$CDM models. The KS+Spinor MCG model yields a minimum chi-square value of $\chi^2_{\rm min} = 1564.39$, which is lower than those of both the curved ($\chi^2 = 1572.05$) and flat ($\chi^2 = 1586.40$) $\Lambda$CDM models. This indicates that the KS+Spinor MCG model provides a better overall fit to the combined data set. The reduced chi-square values for all models are close to unity, suggesting that all models provide statistically acceptable fits. However, the KS+Spinor MCG model achieves a slightly lower value ($\chi^2_\nu = 0.9621$), indicating a marginal improvement.

\begin{table*}[htbp]
\centering
\caption{Statistical comparison of cosmological models using the PP+CC+DESI~DR2+CMB data set. The differences $\Delta \chi^2$, $\Delta \mathrm{AIC}$, and $\Delta \mathrm{BIC}$ are computed relative to the model with the lowest value for each statistic.}
\begin{tabular}{lccccccccc}
\hline\hline
Model & Parameters & $\chi^2$ & $\Delta \chi^2$ & $\chi^2_{\nu}$ & AIC & $\Delta$AIC & BIC & $\Delta$BIC \\
\hline

KS+Spinor MCG 
& 6 
& 1564.39 
& 0 
& 0.9621 
& 1576.39 
& 0 
& 1608.77 
& 14.52 \\

Curved $\Lambda$CDM 
& 3
& 1572.05 
& 7.67 
& 0.9644 
& 1578.05 
& 1.67 
& 1594.25 
& 0 \\

Flat $\Lambda$CDM 
& 2 
& 1586.40 
& 22.01 
& 0.9727 
& 1590.40 
& 14.01 
& 1601.20 
& 6.95 \\

\hline\hline
\end{tabular}
\label{tab:model_comparison}
\end{table*}

\begin{figure}[!ht]
\centering
\includegraphics[width=0.8\textwidth]{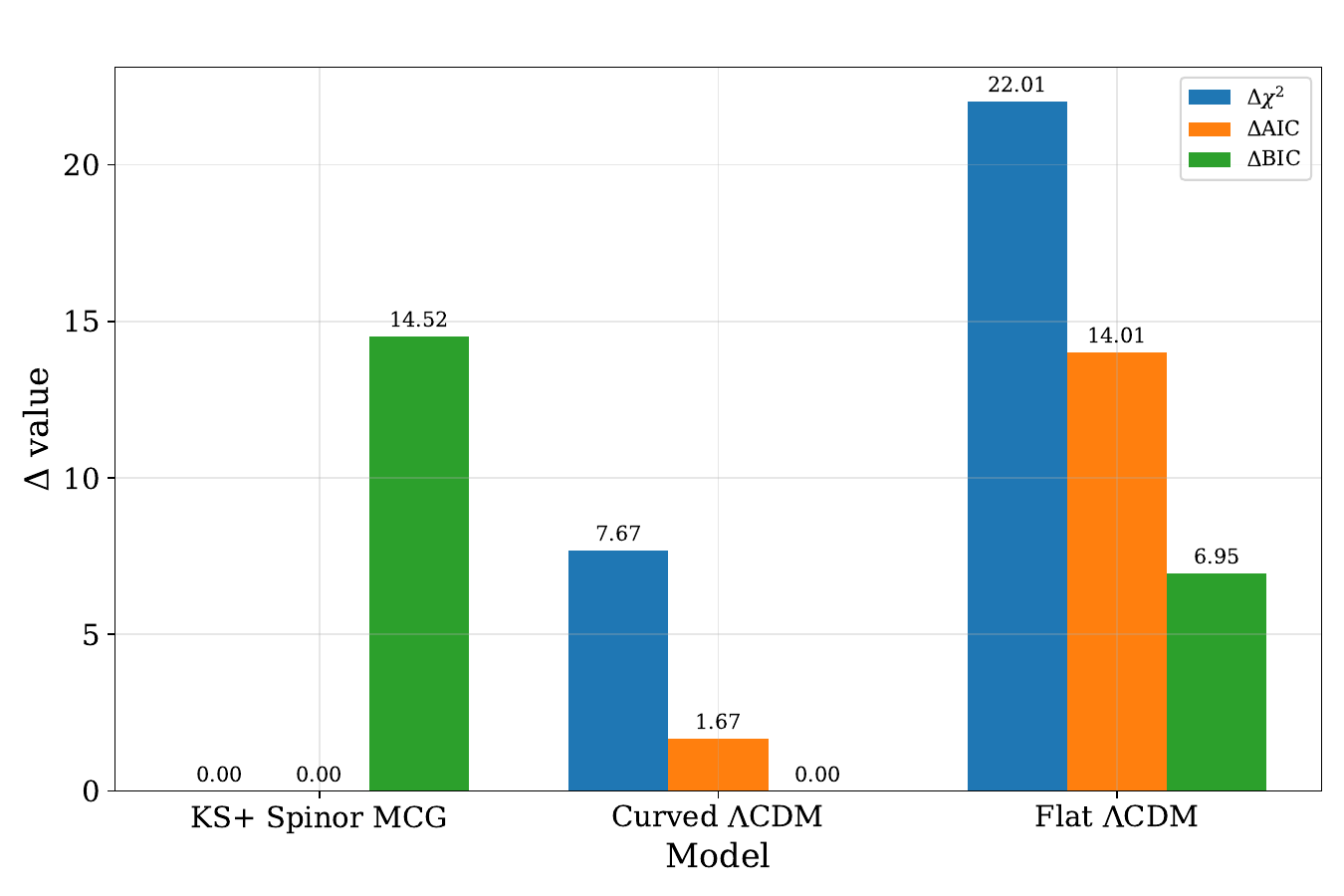}
\caption{Bar plot comparing KS+Spinor MCG, curved $\Lambda$CDM, and flat $\Lambda$CDM models using $\Delta\chi^2$, $\Delta$AIC, and $\Delta$BIC, evaluated relative to the best-fit model for the combined PP+CC+DESI DR2+CMB data set.}
\label{Fig:barplot}
\end{figure}

In terms of information criteria, the KS+Spinor MCG model has the lowest AIC value, confirming that the improved fit justifies the inclusion of additional model parameters. Specifically, $\Delta \mathrm{AIC} = 1.67$ for the curved $\Lambda$CDM model and $\Delta \mathrm{AIC} = 14.01$ for the flat $\Lambda$CDM model, indicating strong evidence against the flat $\Lambda$CDM model. On the other hand, the Bayesian Information Criterion (BIC), which penalizes additional parameters more strongly, favors the curved $\Lambda$CDM model. The KS+Spinor MCG model has $\Delta \mathrm{BIC} = 14.52$ relative to the curved $\Lambda$CDM model, suggesting that while KS+Spinor MCG provides a better fit, the simpler model is statistically preferred under BIC. In Fig.~\ref{Fig:barplot}, we show the statistical comparison between the models using bar plots.

\begin{figure}[!ht]
\centering
\includegraphics[width=0.8\textwidth]{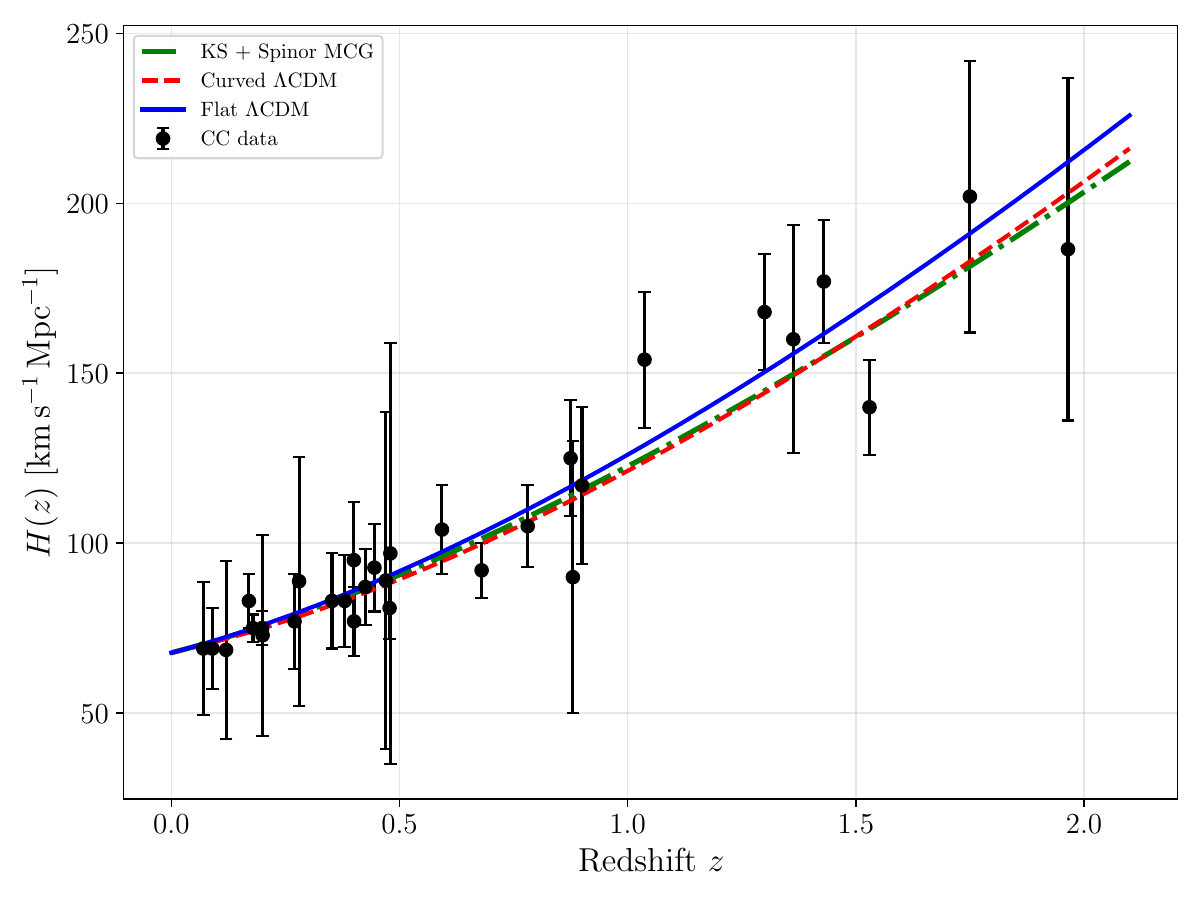}
\caption{Evolution of the Hubble parameter $H(z)$ as a function of redshift $z$. The best-fit curves for the KS+Spinor MCG, curved $\Lambda$CDM, and flat $\Lambda$CDM models are compared with observational cosmic chronometer (CC) data points.}
\label{Fig:Hz_CC}
\end{figure}

\begin{figure}[!ht]
\centering
\includegraphics[width=0.8\textwidth]{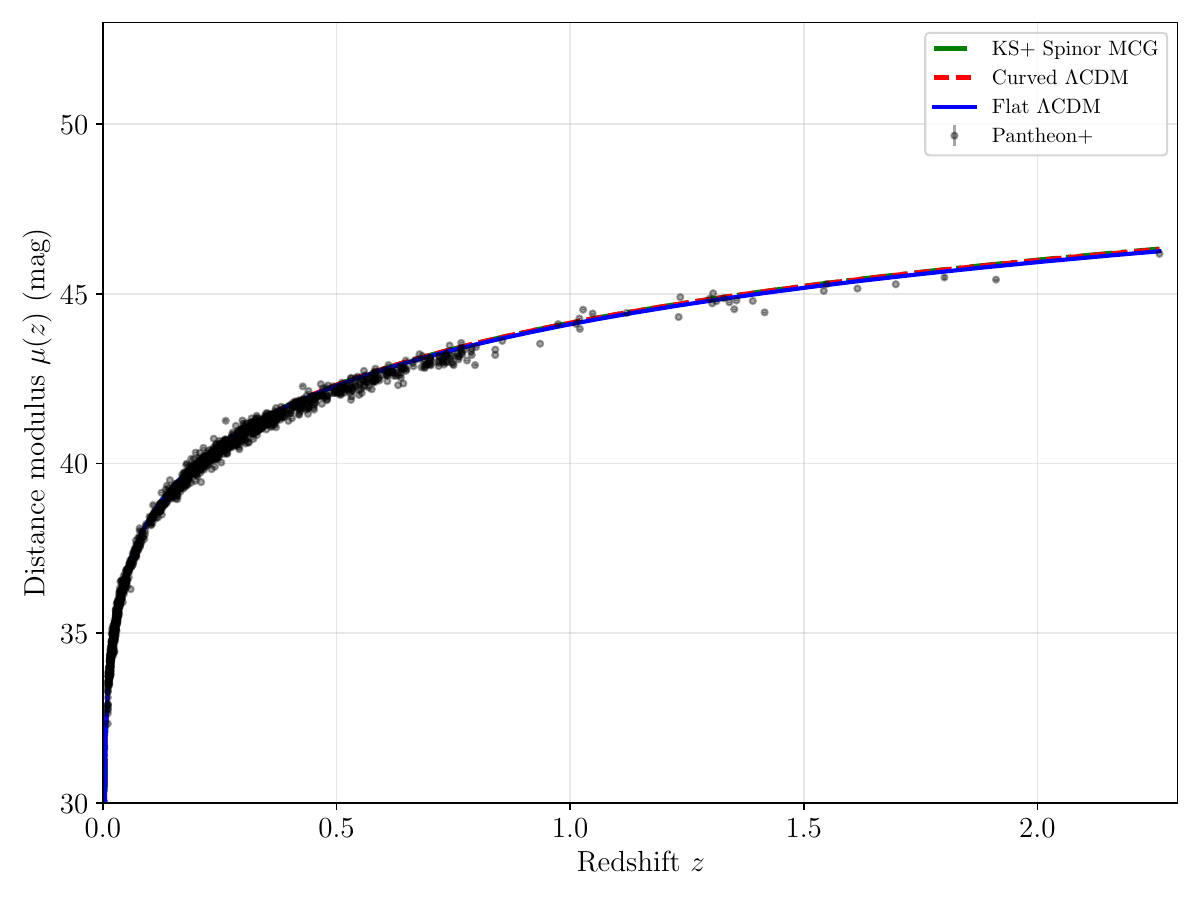}
\caption{Distance modulus $\mu(z)$ versus redshift $z$ for KS+Spinor MCG and $\Lambda$CDM models, compared with Pantheon+ supernova data.}
\label{Fig:mu}
\end{figure}

\begin{figure}[!ht]
\centering
\includegraphics[width=0.8\textwidth]{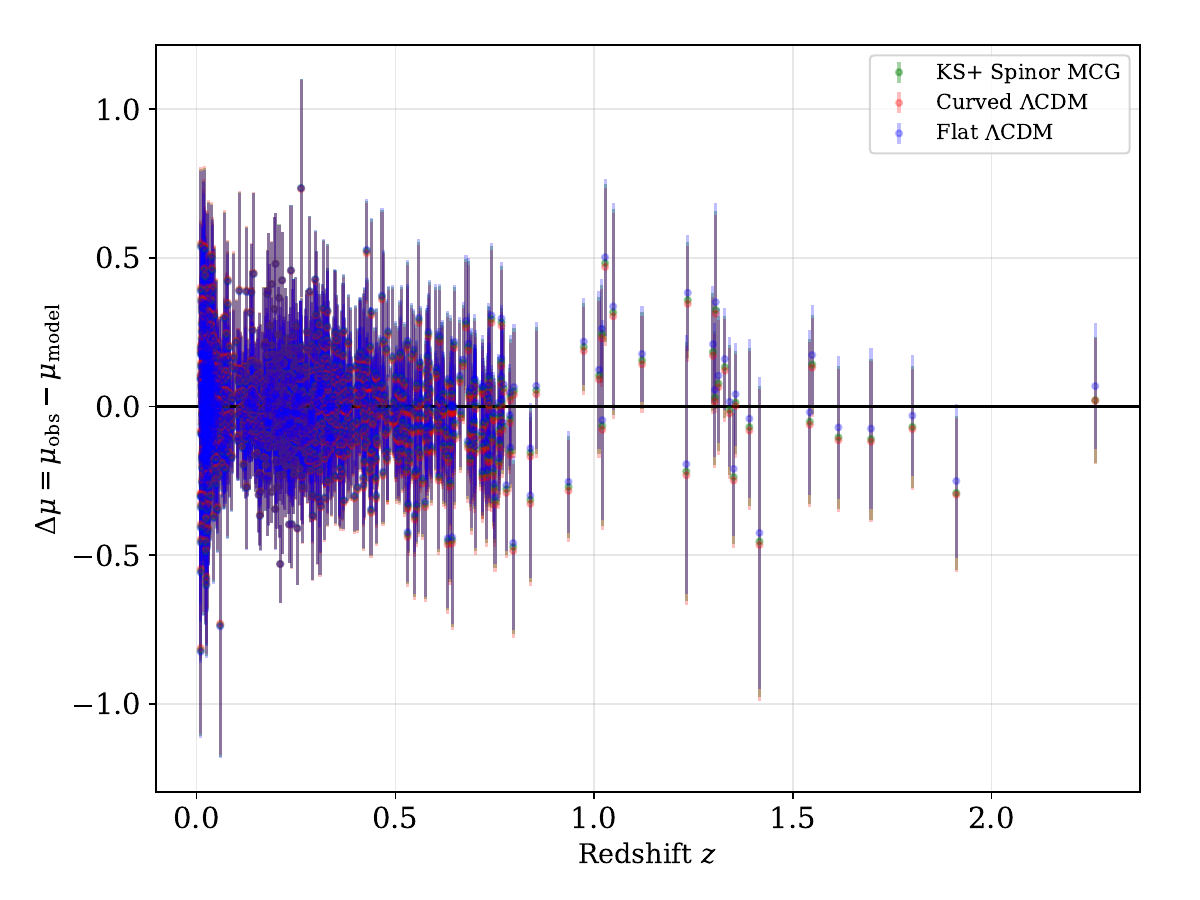}
\caption{Distance modulus residuals $\Delta\mu$ versus redshift $z$ for KS+Spinor MCG and $\Lambda$CDM models using Pantheon+ data.}
\label{Fig:mu_residual}
\end{figure}

Fig.~\ref{Fig:Hz_CC} shows the evolution of the Hubble parameter $H(z)$ with redshift. The best-fit curve is closely aligned with the observational CC data points. The green dash-dotted line (KS+Spinor MCG) closely follows the red dashed line (curved $\Lambda$CDM), while the solid blue line (flat $\Lambda$CDM) shows slight deviations from them. In Fig.~\ref{Fig:mu}, we compare the best-fit distance modulus $(\mu_{\rm best\text{-}fit})$ for the KS+Spinor MCG and $\Lambda$CDM models with the Pantheon+ supernova data from distant galaxies. All models show good agreement with the observational data points. Additionally, we compute the residuals of the distance modulus for different models, as shown in Fig.~\ref{Fig:mu_residual}, where the residual is defined as $\Delta \mu = \mu_{\rm observation} - \mu_{\rm best\text{-}fit}$. Most of the residuals are clustered around zero, indicating that the KS+Spinor MCG model provides a reasonably good fit over the observed redshift range.

\begin{figure}[!ht]
\centering
\includegraphics[width=0.8\textwidth]{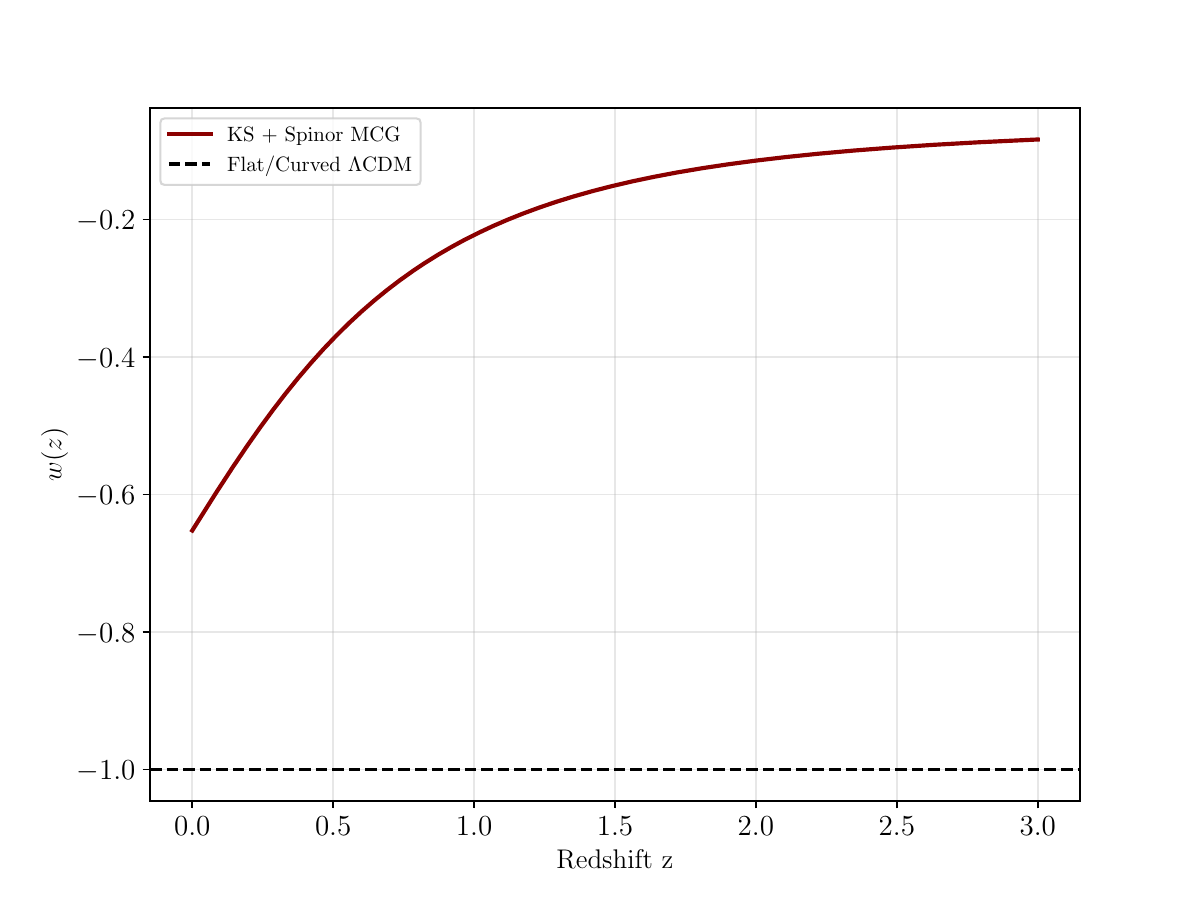}
\caption{Evolution of the effective equation of state parameter $w(z)$ as a function of redshift $z$ for the KS+Spinor MCG model, compared with the constant equation of state $w=-1$ of the $\Lambda$CDM models.}
\label{Fig:wz}
\end{figure}

The dynamical effective equation of state (EOS) parameter of the KS+Spinor MCG model evolves smoothly with redshift, as shown in Fig.~\ref{Fig:wz}. The present value is $w(0) = -0.652$, indicating that the unified fluid behaves like dark energy at late times and approaches pressureless matter at high redshift. This behavior confirms that the spinor field MCG model naturally provides a unified description of dark matter and dark energy within the KS geometry.

\begin{figure}[!ht]
\centering
\includegraphics[width=0.8\textwidth]{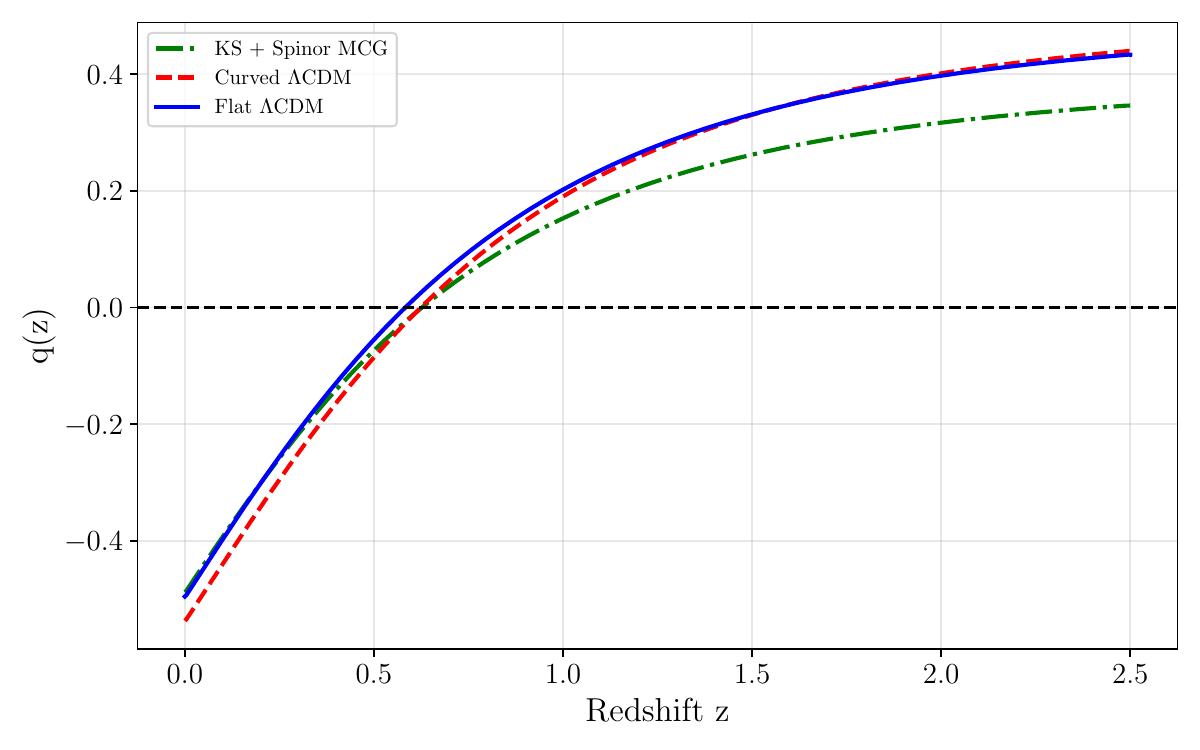}
\caption{Evolution of the deceleration parameter $q(z)$ as a function of redshift $z$ for the KS+Spinor MCG, curved $\Lambda$CDM, and flat $\Lambda$CDM models.}
\label{Fig:qz}
\end{figure}

A clear signature of the accelerated expansion of the Universe is illustrated in Fig.~\ref{Fig:qz}. The present value of the deceleration parameter for the KS+Spinor MCG model is $q(0) = -0.488$, while the curved and flat $\Lambda$CDM models yield $q(0) = -0.537$ and $q(0) = -0.495$, respectively. The evolution of the deceleration parameter in the $\Lambda$CDM models is closely aligned, whereas the KS+Spinor MCG model approaches them from higher redshift and overlaps with the flat $\Lambda$CDM model near the transition redshift $z_t \approx 0.61$.

\section*{VI. Conclusion}

In this work, we investigated a massless nonlinear spinor field based Modified Chaplygin Gas (MCG) dark energy model in Kantowski--Sachs spacetime and constrained its parameters using recent cosmological data sets, namely Pantheon+, CC, DESI DR2, and CMB, across three different data combinations. For comparison, we also analyzed the curved and flat $\Lambda$CDM models using the same observational data.

The MCMC analysis demonstrates that the KS + Spinor MCG model successfully reproduces the observed expansion history of the Universe while incorporating anisotropic effects and a unified dark sector. The shear parameter $\Delta_0$ is found to be consistent with zero within uncertainties, indicating that the present Universe is effectively isotropic. This supports the expectation that anisotropic cosmological models can evolve toward isotropy at late times. Additionally, the curvature parameter approaches values consistent with a spatially flat Universe when CMB data are included, further aligning the model with standard cosmological observations.

The dynamical behavior of the MCG parameters provides a natural unified description of dark matter and dark energy within a single fluid framework, eliminating the need to treat these components separately. In particular, the negative value of the intrinsic equation of state parameter $W$ supports the emergence of late-time cosmic acceleration, while the small value of $\alpha$ indicates only mild deviation from standard Chaplygin gas behavior.

From a statistical perspective, the KS+Spinor MCG model yields a lower minimum $\chi^2$ value compared to both curved and flat $\Lambda$CDM models, indicating a better overall fit to the combined data set. The Akaike Information Criterion (AIC) favors the KS+Spinor MCG model, suggesting that the improved fit justifies the inclusion of additional parameters. However, the Bayesian Information Criterion (BIC), which imposes a stronger penalty for model complexity, prefers the simpler curved $\Lambda$CDM model. This highlights the well-known trade-off between goodness of fit and model simplicity.

Overall, the spinor field MCG model in the Kantowski--Sachs spacetime provides a viable and competitive alternative to the $\Lambda$CDM model. It not only accommodates a unified dark sector but also incorporates anisotropic effects while remaining consistent with current observational constraints. Future high-precision surveys such as Euclid and LSST may further test anisotropic unified dark sector models.\\
\\
\noindent
{\bf Declarations:}

\vskip 5 mm {\bf Competing interests} { There is
no conflict of interests.}


\vskip 5 mm {\bf Funding}  Not applicable.

\vskip 5 mm {\bf Availability of data and materials} No new data sets were generated during the current study.

\normalfont
\end{document}